\documentclass[conference]{IEEEtran}
\IEEEoverridecommandlockouts
\usepackage{cite}

\usepackage{amsmath,amssymb,amsfonts}
\usepackage{algorithm}
\usepackage{algpseudocode}
\usepackage{graphicx}
\usepackage{booktabs}
\usepackage{array}
\usepackage{url}
\usepackage{xcolor}
\usepackage[hidelinks]{hyperref}

\definecolor{ftdgreen}{HTML}{188038}

\begin{document}

\title{Quantization-Tolerant Gradient Auditing for Active Inference-Attack Detection in Federated Learning\\
\thanks{This project is supported by the Federal Ministry for Economic Affairs and Climate Action (BMWK) on the basis of a decision by the German Bundestag.}
}

\author{%
\IEEEauthorblockN{Michael Kuehne-Schlinkert}
\IEEEauthorblockA{\textit{Katulu GmbH}, 
Hamburg, Germany \\
Email: michael@katulu.io}
}

\maketitle

\begin{abstract}
Inside a federated learning system a participant can quietly study the shared model and learn about others' private data. The usual defense, differential privacy, often adds so much noise that the model is no longer worth training. In cross-silo FL, where participants are a few accountable organizations and the server is trusted, we ask whether the server can instead interrogate clients. Each round it sends a secret challenge batch and asks every client to return a checksum of the gradient that batch induces, together with a timing. A client that mutated the model state, as an active generative-adversarial-network discriminator attacker does, returns a wrong checksum. We make this checksum robust to benign cross-platform floating-point non-determinism with a quantized fingerprint compared under a small tolerance. It flags two distinct active attacks at a true-positive rate of $1.0$ with no false positives on the checksum axis, stable across the fingerprint's parameters and up to $50$ clients at malicious ratios to $40\%$, for $32\,\mathrm{ms}$ of server work and $4.35\,\mathrm{kB}$ per response. Against a purely passive attacker the only signal is timing, and it is not robust: a cohort-relative test detects only in low-jitter conditions while a false-positive-free per-client baseline is blind, so no configuration delivers both. An attacker that answers the challenge on an untouched copy of the model while tampering with a replica evades the audit entirely. Therefore the checksum is a reliable in-band signal against attackers that do not decouple the audited computation from the attacked one, and it composes with differential privacy, whose accuracy cost we quantify on Texas100, rather than replacing it.
\end{abstract}


\section{Introduction}
\label{sec:intro}

Federated learning (FL) lets several parties train a shared model without moving raw data off-premises~\cite{mcmahan2017}. In cross-silo settings the parties are a handful of organizations (e.g., manufacturers cooperating on a quality-control model) each holding a large local dataset. The promise is that data never leaves the silo. The reality is more subtle, because the model updates exchanged every round are functions of the local data, and a participant willing to analyze what the protocol already hands it can recover information about the data of others.

Two families of leakage are relevant in this setting. Membership inference asks whether a particular record was in a client's training set~\cite{shokri2017}. Property inference asks about statistical properties of that set (e.g., the proportion of a sensitive attribute)~\cite{melis2019}. Both can be mounted passively. The attacker follows the protocol exactly, sends well-formed updates, and only reads the global models it receives. Nasr et al.\ showed a passive FL participant extracts a strong membership signal from the iterative parameter exchange~\cite{nasr2019}. Because nothing about the attacker's messages looks wrong, the prevailing view is that passive inference cannot be detected inside the unmodified protocol. Leaving differential privacy (DP) ~\cite{abadi2016} as the only recourse, at the cost of the accuracy that motivated forming the federation.

This paper questions that recourse for cross-silo FL, which has a structural property worth exploiting. The server is operated under a contractual and audit framework and is treated as trusted, while clients may be curious or actively malicious. Under that assumption the server is not limited to averaging updates. It can interrogate the participants.

The mechanism we introduce is a challenge-response audit. In each round the server constructs a small challenge batch that is not know to any other party, evaluates the gradient it induces on the current global model, reduces it to a reference checksum, and asks every client to return the same checksum together with the latency of computing it. A client that evaluated the model it was sent reproduces the checksum. A client that mutated it, as an active generative adversarial network (GAN) attacker does when it repurposes the global model as a discriminator~\cite{hitaj2017}, produces a different checksum. A client running an extra analytic membership test instead spends measurable extra time, which surfaces as a latency anomaly.

Our central finding is a split between the two signals, established by measurement. The checksum channel is reliable: with the quantization tolerance it catches the active, model-mutating attacks the challenge gradient can see, at a true-positive rate of $1.0$ and essentially no false positives. It holds across the fingerprint's parameters and up to $50$ clients. The timing channel, the only signal against a passive attacker, is not: a cohort baseline detects such an attacker in clean conditions but false-positives under realistic wide-area jitter, while the false-positive-free per-client baseline is blind to it.  

This paper makes the following contributions:
\begin{itemize}
  \item A \emph{quantization-tolerant challenge-gradient checksum}. That a trusted server can flag a client which mutates the shared model is almost immediate. Making it deployable across heterogeneous silos is not, because an exact gradient hash false-positives honest clients on real heterogeneous hardware. Our tolerant fingerprint absorbs that drift while still catching mutation, detecting two distinct active attacks at a true-positive rate of $1.0$ with no false positives, stable across the fingerprint's parameters, at $K$ up to $50$ clients and malicious ratios up to $40\%$.
  \item A negative result on timing-based detection of \emph{passive} analytic inference: a cohort-relative latency test detects an always-on passive attacker only in low-jitter conditions and false-positives under realistic wide-area jitter, while a false-positive-free per-client baseline is blind to it, so no configuration delivers both. We quantify this tradeoff across the jitter range.
  \item A differential privacy comparison on the public Texas100 benchmark tracing a privacy-utility frontier where tuned centralized DP-SGD either leaves residual leakage at a moderate accuracy cost or closes the channel only by collapsing accuracy about thirty points. DP bounds leakage \emph{before} and the audit detects \emph{after}, so both complement.
  \item A threat-model and cost analysis: the checksum and timing leak no client data to an honest-but-curious server, since the audited gradient is taken on the server's own challenge batch, at $32\,\mathrm{ms}$ of server work per round and, with the challenge-derived block subsampling of Section~\ref{sec:cr}, $4.35\,\mathrm{kB}$ per response. We measure the evasion surface rather than asserting it, including a clean-copy attacker that defeats the audit outright.
\end{itemize}

\section{Background and Related Work}
\label{sec:related}

\subsection{Passive inference in federated learning}

Membership inference against centralized models was formalized by Shokri et al. via shadow models that distinguish a model's behavior on members from non-members~\cite{shokri2017}. In FL the attacker is a participant inside the training loop, not an external querier, which strengthens the attack. Nasr et al. showed that the iterative exchange of model parameters is itself a leakage channel, exploitable by a participant who only observes the global models~\cite{nasr2019}. Melis et al. demonstrated property leakage from collaborative learning, recovering attributes uncorrelated with the task~\cite{melis2019}. FedMIA~\cite{fedmia2025} strengthens the passive attacker further by combining updates from \emph{all} clients across rounds rather than only the target's, sharpening exactly the threat our timing channel fails to catch. Hitaj et al. introduced a generative attack in which the jointly trained model serves as a GAN discriminator~\cite{hitaj2017}. Gradient-inversion attacks reconstruct training samples directly from shared gradients~\cite{zhu2019}, a server-side threat orthogonal to the client-side inference we detect. Bai et al.\ survey this landscape, noting that FL attacks target the training phase throughout convergence rather than a released model~\cite{bai2025survey} -- which is what makes an in-round audit conceivable at all. These attacks share the same characteristic that makes detection difficult. In the passive case the attacker's outbound messages are protocol-conforming, so a server that inspects only update norms or directions observes nothing anomalous. The active variants do deviate, but can be tuned to stay within the tolerances of robust aggregation.

\subsection{Differential privacy and its cost in cross-silo FL}

DP bounds the influence of any single record on the released model~\cite{dwork2006} and is the standard defense against membership inference~\cite{abadi2016}. Its protection is real but purchased with noise. In cross-silo industrial FL the noise needed to close the inference channel often pushes accuracy below the level that justified the federation (Section~\ref{sec:dp}). This motivates a detection-first approach: identify the participant mining the model and respond through governance, rather than blunt every update.

\subsection{Byzantine robustness and the conflation problem}

A parallel line of research hardens aggregation against malicious updates. Krum selects the single update with the smallest summed distance to its nearest neighbors~\cite{blanchard2017}. Coordinate-wise trimmed mean and median bound the influence of extreme updates with provable rates~\cite{yin2018}. Shejwalkar and Houmansadr constructed stronger model-poisoning attacks and a divide-and-conquer defense based on singular-value decomposition of the update matrix~\cite{shejwalkar2021}. These defenses answer the question \emph{is this update corrupting the model}, which is distinct from the question \emph{is this participant reading the model}. An active inference attacker can perturb its updates to amplify leakage while remaining inside robustness tolerances. AgrEvader makes the point explicit, hiding a poisoned update inside a cover set of clean samples so that coordinate-wise and distance-based robust aggregators do not react~\cite{zhang2023}. Separating manipulation from curiosity is an open problem, which Section~\ref{sec:results} addresses.

\subsection{Positioning}

The audit is related to verifiable computation and model watermarking, but narrower and cheaper. It neither certifies a client's training nor embeds ownership marks. It is also stronger than naively attesting the model weights a client claims to hold. The checksum binds to a server-secret challenge batch the client cannot precompute, and it verifies the model's \emph{function} on that batch rather than a self-reported weight hash, which a client keeping a clean copy could forge while tampering with a replica. Combining a server-chosen challenge-gradient checksum with a latency side-channel is, to our knowledge, a novel server-side audit for cross-silo FL, though only the checksum axis proves reliable (Section~\ref{sec:results}).

\section{Threat Model}
\label{sec:threat}

We consider horizontal cross-silo FL with synchronous rounds and weighted averaging in the style of FedAvg~\cite{mcmahan2017}. There are $K$ clients indexed by $c \in \{1,\dots,K\}$, with $K$ on the order of five to fifty named organizations. In round $t$ the server distributes the global model $\theta_t$, each client $c$ returns an update $u_t^{(c)}$, and the server aggregates $\theta_{t+1} = \theta_t + \sum_c \alpha_c\, u_t^{(c)}$ with non-negative weights $\alpha_c$.

\paragraph*{Server} The server is trusted and not corruptible, reflecting the commercial reality of cross-silo deployments under an information-security framework. The server holds a secret seed $s$ that no client observes. A malicious or compromised server is out of scope: the challenge-response channel is itself a probing primitive, so defending clients against a curious server is a separate problem we do not address.

\paragraph*{Clients} Clients fall into two behaviors.
\begin{itemize}
  \item \emph{Honest-but-curious.} The client follows the protocol exactly and attempts to learn from what the protocol exposes. Passive cosine membership inference belongs to this class.
  \item \emph{Malicious.} The client deviates, either to corrupt the model (Byzantine) or to amplify leakage (active inference). An active GAN attack, which mutates the local copy of the global model, and AgrEvader, which perturbs the reported update, belong to this class.
\end{itemize}

\paragraph*{Attacker capabilities} The attacker observes the global model in each round and its own local data, and may run arbitrary local computation. It receives each round's challenge specification but not the server's secret seed, so it cannot predict future or past challenges. The defender's goal is not to prevent a client from reasoning about the models it legitimately receives, which is impossible, but to detect inference at the protocol level so the federation can quarantine, ban, or pursue the participant under contract. Evasion attacks at inference time are out of scope, since trained models are operated locally inside each silo.

\paragraph*{Evasion options} A sophisticated attacker can \emph{decouple} the inference to an offline thread, \emph{pad} its latency to the cohort median, or keep a \emph{clean copy} of the model to answer the challenge while tampering with a replica. The audit targets the thorough in-band attacker. Section~\ref{sec:evasion} measures where each evasion succeeds.

\paragraph*{Honest-but-curious server} The audit hands even a curious server nothing. the challenge gradient is computed on the server's \emph{own} challenge batch, and the client returns a hash of it plus a scalar latency, never a function of the its private data.

\section{Method}
\label{sec:method}

\subsection{Challenge-response gradient auditing}
\label{sec:cr}

\subsubsection{Challenge construction}
At round $t$ the server derives a per-round key from its secret seed $s$ as
\begin{equation}
k_t = \mathrm{SHA\text{-}256}(s \,\|\, t),
\end{equation}
where $\|$ denotes string concatenation. The key $k_t$ seeds a deterministic draw of a challenge batch $B_t$ of fixed size $m$ (we use $m=16$). The batch is reproducible by the server in every round and is unknown to clients.

\subsubsection{Reference checksum}
\label{sec:determinism}
The server evaluates the gradient of the loss $\ell$ on $B_t$ at the current global model,
\begin{equation}
g_t = \nabla_\theta \, \ell(\theta_t; B_t),
\end{equation}
and reduces it to a reference fingerprint $h_t$; we call the comparison a \emph{checksum} for brevity, though it is a tolerance-thresholded fingerprint, not an exact digest. One subtlety decides whether it is deployable at all. Floating-point gradient evaluation is not bitwise reproducible across hardware (non-associative summation, fused multiply-add, ARM-versus-x86, CPU-versus-GPU), so a raw FP32 hash reads benign drift as a deviation. Across four machines at matched software - two Apple-silicon laptops, an x86\_64 Linux server and an arm64 Linux server - the exact whole-gradient hash disagrees in every heterogeneous pair, while the tolerant fingerprint stays below $\tau$ in all of them. The honest mismatch tracks how far apart the two stacks are: $0.0042$ between two machines of the same architecture and operating system, $0.0208$ across operating systems at fixed architecture, $0.0250$ across architectures at fixed operating system, and $0.0321$ when both differ. Instruction set is the largest single term, and the two effects compose sub-additively. The drift is nonetheless \emph{deterministic} and not noisy: two x86\_64 instances on different continents produced a byte-identical challenge gradient, so the mismatch is a property of the stack (architecture, operating system, library) and not of the individual host, and a silo pair can be calibrated once. Divergence is also not guaranteed - under a different library version two of these machines agreed bitwise - which is the stronger argument for a tolerance. An operator cannot know in advance which of its silos will agree, since that depends on a software stack it does not control, so an audit requiring exact agreement is one library upgrade away from false-positiving an honest participant.

The robust fingerprint quantizes each coordinate relative to the gradient's own mean magnitude $q\,\overline{|g_t|}$, splits the result into blocks of fixed size, hashes each, and compares clients on the \emph{fraction of differing blocks} against a tolerance $\tau$ ($q=10^{-2}$, $256$ coordinates per block, $\tau=0.05$). Two properties let it absorb drift while still catching mutation. The relative step is \emph{scale-invariant}: under $g_t\mapsto\alpha g_t$ the step $q\,\overline{|g_t|}$ scales by the same $\alpha$, so every quantized coordinate, and hence every block digest, is unchanged. A global rescaling of an honest gradient, what cross-process reduction-order differences mostly produce, leaves the fingerprint untouched. And fixing block \emph{size} instead of block count keeps the honest mismatch bounded independent of model size, where a fixed count saturates on large models. Against an honest worst case of $0.0321$ over the pairs we measured, a model-mutating GAN attack moves about $18\%$ of coordinates and $0.89$ of blocks: $\tau=0.05$ clears the honest maximum by $1.6\times$ and sits $17.7\times$ below the attack floor.

GPU evaluation is the large exception. Its divergence from CPU is input-dependent, ranging in our measurements from $0.016$ to $0.868$ -- at the top of that range indistinguishable from the mutation's $0.89$, so the separation does not survive admitting it. Therefore, the protocol \emph{mandates} CPU evaluation of the challenge, which costs a client one forward-backward pass over $16$ samples. This requirements also solves a problem the the server would otherwise arbitrate: if a client returns a mismatched fingerprint and blames its GPU, which has not comply with the protocol, and the audit treats that as tampering.

\subsubsection{Block subsampling}
As described, the fingerprint carries one digest per block, so it grows with the model: $102{,}601$ blocks, or $1.74\,\mathrm{MB}$, for the $26.3$M-parameter network of Section~\ref{sec:setup}. Because $k_t$ is already secret, the server can derive from it a random subset of $S$ blocks and require only those, bounding the response \emph{and} the hashing cost independently of model size. The subset is unpredictable for the same reason the challenge batch is, so a client can neither precompute its answer nor choose which coordinates are safe to tamper with. A subsample of $S$ misses a mutation moving a fraction $\phi$ of blocks with probability $(1-\phi)^S$: at the measured $\phi=0.89$ and $S=256$, $3\times10^{-253}$, for a $4.35\,\mathrm{kB}$ response.

\subsubsection{Client response}
Alongside its regular update $u_t^{(c)}$, each client returns the checksum $h_t^{(c)} = H\!\left(\nabla_\theta \ell(\hat\theta_t^{(c)}; B_t)\right)$ of its challenge gradient, computed on the model state $\hat\theta_t^{(c)}$ that the client actually used, and a timing $\lambda_t^{(c)}$: the server-observable wall-clock latency of the challenge computation, which is the timing signal we use throughout. A thread-scoped work clock (per-thread CPU time) would strip out the network and co-tenant noise that limits the wall clock, but it is client-reported and therefore forgeable by the very adversary we target.

\subsubsection{Indicator 1, checksum deviation}
A returned fingerprint whose block-mismatch fraction against the reference exceeds the tolerance $\tau$,
\begin{equation}
d_t^{(c)} = \mathbb{1}\!\left[ \mathrm{mismatch}\!\left(h_t^{(c)}, h_t\right) > \tau \right],
\end{equation}
is hard evidence that the client evaluated the challenge on a model state $\hat\theta_t^{(c)} \neq \theta_t$. An active GAN attacker, which mutates its local model to use it as a discriminator, triggers this indicator: it perturbs about $18\%$ of coordinates, so essentially every block differs and the mismatch fraction sits far above $\tau=0.05$. The tolerance is what makes the test deployable: an exact hash comparison, with $\tau=0$, false-positives on honest clients, because benign floating-point non-determinism perturbs coordinates and flips the digest even though the client reproduced $\theta_t$. The quantized comparison absorbs that drift, driving the honest block-mismatch fraction to zero, so the false-positive rate from this channel is zero in our measurements.

\subsubsection{Indicator 2, latency anomaly}
Local round latencies are scored per round with a modified z-score based on the median and the median absolute deviation (MAD),
\begin{equation}
z_t^{(c)} = \frac{0.6745 \left( \lambda_t^{(c)} - \widetilde{\lambda}_t \right)}{\mathrm{MAD}_t},
\label{eq:modz}
\end{equation}
where $\widetilde{\lambda}_t$ is the median latency in round-$t$ and the constant $0.6745$ rescales the MAD to the standard deviation under normality. To avoid division by a near-zero spread the MAD is floored at $\max(\widehat{\mathrm{MAD}}_t,\ \rho|\widetilde\lambda_t|,\ \lambda_{\min})$ with $\rho=0.1$ and $\lambda_{\min}=1\,\mathrm{ms}$. The relative floor $\rho\,|\widetilde\lambda_t|$ dominates for deep networks whose rounds take seconds, the absolute one only for sub-second tabular rounds, so the detector's sensitivity tracks round duration. A client is flagged in round $t$ when $z_t^{(c)}>\kappa$. This threshold trades detection against false alarms and is the detector's most important parameter: we report results at the benchmark's $\kappa=2.5$, at the library default $\kappa=5.0$, and at an extreme-value $\kappa(K)$ in Section~\ref{sec:scale}. A per-client variant scores each client against its own historical median and MAD with a two-round warm-up, for federations whose clients have heterogeneous but individually stable hardware.

\subsubsection{Indicator 3, the combined feature}
Per client the server forms a two-dimensional feature from the run-level latency-flag count $L^{(c)} = \sum_t \mathbb{1}[z_t^{(c)} > \kappa]$ and the run-level checksum-deviation indicator $D^{(c)} = \max_t d_t^{(c)}$, and classifies according to
\begin{equation}
\mathrm{role}(c) =
\begin{cases}
\text{inference} & L^{(c)} \ge L_{\min} \\[2pt]
\text{Byzantine} & L^{(c)} < L_{\min} \text{ and } D^{(c)} = 1 \\[2pt]
\text{honest} & \text{otherwise,}
\end{cases}
\label{eq:role}
\end{equation}
with persistence threshold $L_{\min} = 2$ for runs of more than one round. The rationale follows the attack physics. A persistent latency anomaly without model-state corruption is consitent with a passive analytic membership test that costs time but does not alter the model, which maps to inference. A checksum deviation without persistent latency is a client corrupting the model state for reasons unrelated to reading it, which maps to Byzantine. A generative attacker whould trip both signals, but in practice its extra compute stays below the latency threshold (Section~\ref{sec:results}), so it trips the checksum alone and is grouped with state mutation. This is not a edge case: of the $16$ caught active-GAN instances, $15$ are labelled Byzantine instead of inference, so the active inference attacker is systematically classified as manipulation -- the very conflation the rule was meant to resolve. Therefore, the equation~\eqref{eq:role} has one working axis. Its checksum axis separates state mutation from honest behavior at the rates Table~\ref{tab:cr} reports, while its inference axis is fed by the timing channel and inherits that channel's fragility (Section~\ref{sec:jitter}), and does not reliably separate curiosity from manipulation. Algorithm~\ref{alg:cr} summarizes the round-level procedure.

\begin{algorithm}[t]
\caption{Challenge-response audit, server side, round $t$}
\label{alg:cr}
\begin{algorithmic}[1]
\State $k_t \gets \mathrm{SHA\text{-}256}(s \,\|\, t)$
\State $B_t \gets \textsc{DeriveBatch}(k_t, m)$
\State $g_t \gets \nabla_\theta \ell(\theta_t; B_t)$,\quad $h_t \gets H(g_t)$
\State broadcast $\theta_t$ and the challenge specification
\For{each client $c$}
  \State receive $u_t^{(c)},\, h_t^{(c)},\, \lambda_t^{(c)}$
  \State $d_t^{(c)} \gets \mathbb{1}[\mathrm{mismatch}(h_t^{(c)}, h_t) > \tau]$
\EndFor
\State compute $z_t^{(c)}$ for all $c$ via \eqref{eq:modz}
\State update $L^{(c)}, D^{(c)}$ and classify roles via \eqref{eq:role}
\State aggregate $u_t^{(c)}$ over clients with $\mathrm{role}(c)=\text{honest}$
\end{algorithmic}
\end{algorithm}

\subsection{Attacker models}
\label{sec:attackers}

Four attacks instantiate the threat model.
\begin{itemize}
  \item \emph{Cosine membership inference}, a passive analytic attack. For each query point the attacker score the cosine similarity of a single-sample gradient against the previous aggregate, mapped to a membership probability $(\cos+1)/2$. It does not mutate the model, so it presents only a latency signal.
  \item \emph{Active GAN attack}, an active generative attack. The attacker optimizes latent inputs by stochastic gradient descent (SGD) over five steps to maximize model confidence on a target class, mutating the local model state in the process, so it presents a checksum deviation reinforced by latency.
  \item \emph{Active gradient-ascent amplification}, a second active attack in the style of the Nasr et al.\ active white-box attacker~\cite{nasr2019}. It runs five gradient-ascent steps on a probe batch to sharpen the model around its data and magnify the membership signal it reads, mutating the model state, so it too presents a checksum deviation.
  \item \emph{AgrEvader-style gradient-orthogonal perturbation}~\cite{zhang2023}, an active inference stand-in. AgrEvader proper conceals its poisoned update within a clean cover set. We use a gradient-orthogonal variant that evades cosine-similarity filters. A random perturbation is orthogonalized against the previous aggregate by Gram-Schmidt and scaled to $0.1$ of the update norm, so it is orthogonal to the aggregate by construction and does not alter the audited challenge gradient.
\end{itemize}

\section{Experimental Setup}
\label{sec:setup}

\paragraph*{Federation} Five clients, synchronous weighted averaging, one or two of them attackers. The detection matrices are run over eight seeds, $\{100,\dots,107\}$; the remaining studies use three or four seeds per configuration. The simulator runs each client as a separate process, which is the source of the benign per-process timing variance of Section~\ref{sec:scale}.

\paragraph*{Primary data and model (public)} CIFAR-10, $1{,}500$ member images sharded over the five clients ($279$ to $317$ each), twelve rounds, on a $2.2$M-parameter convolutional neural network (CNN). This is the setting of Table~\ref{tab:cr} and is fully reproducible: the data is public and the code is released.\footnote{\url{https://github.com/katulu-io/quantization-tolerant-gradient-auditing}} The Texas100 study of Section~\ref{sec:dp} uses the same public benchmark principle for the differential privacy comparison.

\paragraph*{Industrial deployment (non-public)} A cross-silo quality-control dataset -- binary defect classification on $14$ tabular features -- sharded over five silos, thirty rounds, SGD at learning rate $0.0005$ and batch size $32$. The majority (defect-free) class is capped at $200$ records per client and the defect class oversampled, giving shards of $356$ to $414$ records and $1{,}950$ samples in total, against a five-hidden-layer network of width $2560$: $26{,}265{,}601$ parameters, four orders of magnitude more than data. We state that ratio prominently because it bounds what this setting can show. The network is not learning a generalizable defect classifier and none of our claims require it to, since the audit asks whether a client evaluated the model it was sent -- a question about model state, not generalization, and one the challenge gradient answers on the server's own batch rather than on client data. The dataset is the intellectual property of the project partners and cannot be released, which is why it supports the result and not carries it.

\paragraph*{Attacks and detectors} Each attack of Section~\ref{sec:attackers} is evaluated against the challenge-response audit, plus an honest-only baseline for the false-positive rate.

\paragraph*{Metrics} For each configuration the detected set is compared to the known attacker set for the true- and false-positive rates (TPR/FPR). Detector cost per round is recorded separately from total runtime.

\paragraph*{Reproducibility and configurations} The audit uses the per-client latency baseline (two-round warm-up), $\kappa=2.5$, $L_{\min}=2$, and a fingerprint at $q=10^{-2}$, $256$ coordinates per block, $\tau=0.05$. The second active attack runs five gradient-ascent steps on a probe batch before the challenge. The determinism measurement compares the exact hash against the tolerant fingerprint across CPU thread counts, GPU, and separate physical machines (Section~\ref{sec:determinism}). The jitter frontier (Fig.~\ref{fig:frontier}) replays both baselines over eight seeds, adding modeled log-normal jitter at $\{0,\dots,400\}\,\mathrm{ms}$ to the measured $854\,\mathrm{ms}$ industrial round. The short-round comparison uses the $112\,\mathrm{ms}$ CIFAR-10 rounds. The Texas100 study trains a two-hidden-layer MLP (widths $256$, $128$) on $10{,}000$ members, batch $512$, clipping norm $1.0$ -- non-private at $60$ epochs, DP-SGD tuned over learning rate $\{0.5,1,2\}$ and $\{20,40\}$ epochs (Opacus, $\delta=10^{-5}$), best per $\varepsilon$ over three seeds.

\section{Results}
\label{sec:results}

The central result is a split between the two audit signals (Table~\ref{tab:cr}): the checksum reliably detects the active attacks, while the latency channel neither catches the passive attacker nor stays reliably free of false positives.

\subsection{The checksum channel: reliable on active attacks}

We lead with public data. On real CIFAR-10, both active attacks are caught in every seed and every configuration -- a TPR of $1.00$, $48/48$ attacker-instances -- with no false positives anywhere in the matrix (Table~\ref{tab:cr}). We report the separation as a margin rather than as an interval over seeds, because it is not a sampling question: across every configuration the honest block-mismatch fraction is $0.000$ over $320$ honest client-instances, and the smallest of $48$ in-band attacker mismatches is $0.65$, against $\tau=0.05$. The gap exceeds an order of magnitude on both sides and is deterministic. The checksum is a function of model state, so it repeats exactly.

The result is not specific to one attack, architecture, or dataset: the gradient-ascent amplifier is caught identically, the industrial margin below is wider still ($0.000$ against $0.90$), and a compact $0.27$M-parameter network leaves it unchanged (Section~\ref{sec:scale}). The attacker side moves with architecture ($0.65$ convolutional, $0.90$ tabular) while the in-federation honest side stays at $0.000$, which is the useful observation. The binding constraint on $\tau$ is not this matrix but cross-hardware drift (Section~\ref{sec:determinism}), whose worst measured pair reaches $0.0321$.

Equation~\eqref{eq:role} maps the clean-latency, broken-checksum pattern to a Byzantine client, which as Section~\ref{sec:cr} notes mislabels an inference attacker. The passive threat here is real. LiRA~\cite{carlini2022} reaches a $15.8\%$ membership TPR at $0.1\%$ false positives. Since it is passive, it never touches the challenge gradient, so it falls to the timing channel, which, as we show, proves to be not robust.

\paragraph*{Industrial deployment} The same audit runs on a cross-silo quality-control federation at a project partner (five silos, thirty rounds, eight seeds, the $26.3$M-parameter network of Section~\ref{sec:setup}). Both active attacks are caught at a TPR of $1.00$ ($48/48$), honest mismatch $0.000$ against an attacker minimum of $0.90$. Its one false positive -- a single honest instance in $24$, on the latency axis, never the checksum -- is not a stable quantity. Section~\ref{sec:scale} reports what happened when we tried to reproduce it, which motivated the work. Section~\ref{sec:setup} is explicit about what its scale does.

\subsection{The timing channel is not robust}
\label{sec:jitter}
The passive cosine membership test is the case the timing channel exists to catch, and the false-positive-free per-client baseline never catches it: a TPR of $0.00$ in every seed and configuration of Table~\ref{tab:cr}. The analytic test adds a small constant per-round cost, and a baseline that learns each client's own timing absorbs an always-on overhead into the baseline itself, leaving no within-client anomaly to flag. Absorbing benign heterogeneity and absorbing the attack are the same operation.

\begin{figure}[t]
\centering
\includegraphics[width=0.72\columnwidth]{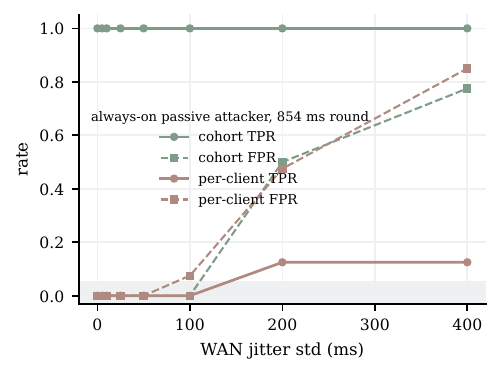}
\caption{The timing channel is not robust. TPR and FPR of the cohort and per-client latency baselines against the always-on passive attacker, under modeled WAN jitter, eight seeds. The cohort baseline detects (TPR $1.0$) but its FPR explodes past $100\,\mathrm{ms}$ of jitter. The per-client baseline is false-positive-free only while blind (TPR $\approx 0$). Neither holds low false positives and high detection together across the modeled jitter range (shaded: FPR $\le 0.05$). The two inter-site links we measured (Section~\ref{sec:jitter}) have a MAD near $2.5\,\mathrm{ms}$.}
\label{fig:frontier}
\end{figure}

The cohort-relative baseline escapes that dilemma, and in clean, homogeneous conditions detects the attacker at a TPR of $1.0$ with no false positives, but only by reintroducing the cross-client comparison benign variance trips. When re-running the detector on the measured latencies with modeled wide-area jitter (Fig.~\ref{fig:frontier}), its honest false-positive rate is zero to $100\,\mathrm{ms}$, $0.50$ at $200\,\mathrm{ms}$ and $0.78$ at $400\,\mathrm{ms}$. The per-client baseline is no fallback, reaching $0.48$ itself at $200\,\mathrm{ms}$ while staying blind. 

The detection also requires the honest round to dominate the jitter: on the $112\,\mathrm{ms}$ CIFAR-10 rounds the cohort false-positive rate passes $0.65$ at $50$--$200\,\mathrm{ms}$, against a negligible rate on the second-scale tabular round. This jitter is modeled, injected onto single-host latencies. Therefore, we measured two real inter-site links from the same client, $300$ TCP samples each: a European-to-North-American link at a median of $109\,\mathrm{ms}$ and a European-to-Australian one at $310\,\mathrm{ms}$. Both are high-latency and stable, with median absolute deviations of $2.4$ and $2.5\,\mathrm{ms}$. Because the detector consumes the MAD and not the absolute latency -- a constant offset is absorbed by the per-round median -- both links sit at the benign end of Fig.~\ref{fig:frontier}. Therefore, the jitter range that breaks the cohort baseline is not where cross-silo links typically occur. The non-reproducibility of Section~\ref{sec:scale} does not depend on the network at all: on one idle host, with no wide-area link involved, identical configurations give honest false-positive rates from $0/16$ to $4/16$. This is a robust reason to distrust the timing axis.

AgrEvader is omitted from Table~\ref{tab:cr} because it evades both channels by construction: its gradient-orthogonal perturbation leaves the challenge gradient unchanged and adds no measurable latency, which we treat as a stated limit, not a measured rate.

\begin{table}[t]
\centering
\caption{Challenge-response detection on \textbf{CIFAR-10}, five clients, twelve rounds, eight seeds, from a single run of the $2.2$M-parameter CNN. Honest block mismatch is $0.000$ across all $320$ honest client-instances against a smallest in-band attacker mismatch of $0.65$, so the checksum column is exactly reproducible and carries no false positives. The clean-copy rows are the adaptive baseline: the same two attacks, answering the challenge on an untouched replica.}
\label{tab:cr}
\begin{tabular}{llcc}
\toprule
Attack (channel) & Atk & TPR & FPR \\
\midrule
honest baseline & $0$ & -- & $0.00$ \\
cosine MI (latency) & $1$ & $0.00$ & $0.00$ \\
cosine MI (latency) & $2$ & $0.00$ & $0.00$ \\
active GAN (checksum) & $1$ & $1.00$ & $0.00$ \\
active GAN (checksum) & $2$ & $1.00$ & $0.00$ \\
active ascent (checksum) & $1$ & $1.00$ & $0.00$ \\
active ascent (checksum) & $2$ & $1.00$ & $0.00$ \\
\midrule
\multicolumn{4}{l}{\emph{adaptive evasion (Section~\ref{sec:evasion})}} \\
GAN, clean copy & $1,2$ & $0.00$ & $0.00$ \\
ascent, clean copy & $1,2$ & $0.00$ & $0.00$ \\
\bottomrule
\end{tabular}
\end{table}

\subsection{Detector cost}

The server's per-round cost is dominated by a part that is easy to remove. Evaluating the reference gradient on the $16$-sample challenge batch takes $25\,\mathrm{ms}$ for the $26.3$M-parameter network of Table~\ref{tab:cr}. Reducing it to a fingerprint costs further $206\,\mathrm{ms}$, because it hashes all $102{,}601$ blocks. Hashing and quantizing only the $S=256$ challenge-derived blocks instead costs $6.8\,\mathrm{ms}$, cutting the total from $231$ to $32\,\mathrm{ms}$ -- $3.8\%$ of the measured $854\,\mathrm{ms}$ honest round rather than $27\%$ -- and the response from $1.74\,\mathrm{MB}$ to $4.35\,\mathrm{kB}$. Scoring all clients is negligible ($0.08\,\mathrm{ms}$ at five clients, $0.79\,\mathrm{ms}$ at fifty). The same audit costs $39\,\mathrm{ms}$ unsubsampled on the $2.2$M-parameter CNN. These costs are carried once per round by the server regardless of cohort size, where aggregation is linear to it.

\subsection{Scale, ratio, and what does not reproduce}
\label{sec:scale}
Three objections concern whether the result survives outside its reported configuration: a five-client cohort, one parameter triple, and an over-parameterized model. Answering them turned up a fourth issue.

\paragraph*{Scale and malicious ratio} We run the real federation at $K\in\{5,20,50\}$ clients with malicious ratios of $10\%$, $20\%$ and $40\%$. The checksum axis is indifferent to both: attacker TPR $1.00$ and honest block mismatch $0.000$ in all eighteen cells, including a $40\%$ malicious cohort. This is structural: each client is compared against the server's own reference, never against its peers, so no number of colluding clients can move the decision boundary -- a property Byzantine-robust aggregation, a statistic over the cohort, cannot have. Sweeping the latency threshold over $\kappa\in\{2.5,\ \max(2.5,\sqrt{2\ln K}),\ 5.0\}$ adds $36$ further cells, and the checksum TPR is $1.00$ in every one: the working axis does not depend on timing parameters.

\paragraph*{Fingerprint parameters and model size} Sweeping the quantization resolution $q$ over $[10^{-4},1]$ and the block size $\{64,\dots,4096\}$ gives $42$ configurations. The honest maximum remains below the attack minimum in all $42$, and $\tau=0.05$ separates them in all $42$: a wide corridor, not a tuned point. A compact $0.27$M-parameter model on identical data gives the same detection as the $26.3$M-parameter one, so the result is not an artifact of over-parameterization.

\paragraph*{The latency axis does not reproduce} Every false positive we observe comes from the latency axis, and its rate rises with cohort size -- at fifty clients and a fixed $\kappa=2.5$ it reaches the tens of percent, as more clients mean more chances for a benign spike to cross the threshold twice. But the more useful finding is that these rates are not stable quantities. Six repeats of each cell, at identical seeds and byte-identical parameters on an idle host, move the honest latency FPR every time: $0/16$ to $4/16$ at $K=5$ and $6/64$ to $18/64$ at $K=20$ on the per-client baseline, $3/64$ to $8/64$ at $K=20$ on the cohort. The checksum was \emph{bit-identical} in all $24$ runs, at TPR $1.00$ and honest mismatch $0.000$. The asymmetry is structural: the checksum is a deterministic function of model state, while the latency axis consumes wall-clock time, and is impacted by all host's activities. The lack of a latency FPR, including the $0.04$ in the industrial deployment, is a property of the detector. It is one sample drawn from a load-dependent distribution, and we report such rates as ranges. This sharpens Section~\ref{sec:jitter}: the timing channel is not just fragile under wide-area jitter, it is not reproducible under identical conditions on one host. Our simulator runs all $K$ clients per host, so co-tenancy increases the magnitudes, but not the non-reproducibility.

\subsection{The evasion surface}
\label{sec:evasion}
Four attacks pass the audit, and we measure each rather than assume it. Gradient-orthogonal perturbation (AgrEvader) leaves the challenge gradient intact by construction. A \emph{timing-padding} attacker that delays its response to the per-round median is reclassified as honest when we replay the detector. An \emph{offline-decoupled} attacker returns a benign update and checksum in-round and mines the saved models on a separate thread. The fourth attack matters most, because it costs the adversary one deep copy. A \emph{clean-copy} attacker answers the challenge on the untouched $\theta_t$ while running its extraction on a replica. Implemented against both active attacks of Table~\ref{tab:cr}, it is never caught on either axis -- a true-positive rate of $0.00$ ($0/48$ attacker-instances over eight seeds), against $1.00$ for the identical attacks run in band. Its fingerprint is not just inside the tolerance but \emph{bit-identical} to an honest client's, block-mismatch $0.000$ in all $72$ instances, since it answers on the untouched $\theta_t$. No choice of $\tau$ can separate it: this is a limit of in-band auditing, not of this fingerprint, since an adversary whose \emph{reported} computation is honest is protocol-indistinguishable from an honest client, whatever the server hashes. What the audit raises is the engineering cost of inference: standard membership and inversion frameworks run inside the synchronous loop, so evading costs an attacker its off-the-shelf tooling.

\subsection{Differential privacy at matched accuracy}
\label{sec:dp}
We test the recurring objection that differential privacy already solves inference, on Texas100~\cite{shokri2017}, a standard membership-inference benchmark ($67{,}330$ records, $100$ classes) that leaks strongly. Two limits bound what this comparison establishes. It is \emph{centralized} per-example DP-SGD while the audit runs in federated cross-silo, because the industrial data cannot carry it: client-level DP-FedAvg noise diverges training at five silos. And the two mechanisms answer different questions -- DP bounds leakage \emph{ex ante} for every attacker, the audit detects \emph{ex post}, after the attacker has read every model it received. Therefore, the comparison is evidence for our \emph{premise}, that the accuracy price of closing the channel with noise motivates looking for a complement, not a claim that the audit beats DP. A multilayer perceptron (MLP) is trained on $10{,}000$ members, non-privately and with per-example DP-SGD (Opacus~\cite{opacus}, formal $(\varepsilon,\delta)$ accounting) tuned over learning rate and epoch budget. We record held-out accuracy and a loss-based membership-inference AUC (MI-AUC)~\cite{yeom2018}. Averaged over three seeds (two for the non-private row), the non-private model leaks clearly: MI-AUC $0.79$ at $53\%$ held-out accuracy. A loose $\varepsilon=8$ holds accuracy at $42\%$ but only lowers the membership signal to $0.55$, still above the $0.5$ chance level. Driving it toward chance ($0.52$) requires $\varepsilon=1$, which collapses accuracy to $21\%$ -- about a thirty-point loss. The audit offers no protection against a purely passive attacker, for which DP remains the only tool that bounds leakage -- at the accuracy cost quantified above.

\section{Discussion and Limitations}
\label{sec:discussion}

\paragraph*{What the checksum does and does not catch} A generative attack moves about $18\%$ of the quantized merely coordinates, which is enough that essentially every block differs, and the checksum flags it cleanly. A subtle attacker could keep its modification below $\tau$ -- a property-inference or targeted minority-membership attack confined to a few channels -- which would fall instead to the timing channel, with the limits we report. The checksum is computed before any communication-efficiency layer: when the platform sparsifies or quantises gradients (Top-$k$, QSGD, signSGD), the server applies the same mask to its reference $g_t$, so honest clients still match by construction.

\paragraph*{Adaptive checksum forgery} A white-box attacker holding $\theta_t$ and the challenge batch $B_t$ can compute $g_t$ itself and pose the GAN attack as a constrained objective, mutating the model while pinning its $B_t$-gradient to $g_t$. This is the gradient-orthogonal evasion of Section~\ref{sec:evasion} applied to the checksum: the mutation must lie in the negligibly thin null space of the challenge gradient. Sweeping a constraint weight $\beta$ that trades attack utility against a null-space penalty (controlled feed-forward model, eight seeds; null-space directional sensitivity about $24\times$), retaining even $1.9\%$ of the unconstrained attack utility already moves $74\%$ of the quantized coordinates, and $4.8\%$ utility moves $78\%$ -- each more than an order of magnitude above $\tau=0.05$. The only checksum-passing point is the unmutated model at zero utility, so forging caps extraction at the honest baseline. The sweep doubles as a utility-versus-detection curve for the in-band attacks of Table~\ref{tab:cr}: $\beta=0$ is the unconstrained GAN at full utility and $0.82$ mismatch, and every point retaining any utility is flagged.

\paragraph*{Detecting gradient-orthogonal attacks} The audit verifies the challenge gradient, not the reported update, so an AgrEvader-style orthogonal perturbation passes. Closing this gap means auditing the update itself: a lightweight route tracks a per-client residual between each reported update and the cohort's descent direction, where a once-per-round orthogonalization leaves a persistent off-trajectory component. A cryptographic route binds the update to honest SGD through trusted-execution attestation or a zero-knowledge proof of training, as TMT-FL~\cite{tmtfl2025} does with a zk-SNARK over the training execution. This is also the only route that closes the clean-copy evasion of Section~\ref{sec:evasion}, because it constrains what the client actually computed. Both cost more than the present hash and timing, so we leave them to future work.

\paragraph*{The timing channel's structural limit} The weakness of Section~\ref{sec:jitter} is structural, not a tuning failure. It is compounded by a reproducibility problem (Section~\ref{sec:scale}). A robust timing detector would need a signal separating inference work from slow hardware, such as an attested per-round work measurement, which we leave to future work.

\paragraph*{Differential privacy and remaining scope} Section~\ref{sec:dp} quantifies on Texas100 the privacy-utility cost the audit avoids. On the proprietary industrial data the comparison is degenerate: client-level DP-FedAvg noise diverges training at five silos, while the compact model shows no membership signal to close.

\section{Conclusion}
\label{sec:conclusion}

A server that interrogates its clients recovers a reliable in-band signal against model-mutating active attacks: a TPR of $1.0$ at negligible cost, no accuracy sacrificed to privacy noise, stable across the fingerprint's parameters and up to $50$ clients. The enabling detail is the quantization-tolerant fingerprint, which absorbs the cross-hardware non-determinism that defeats an exact hash -- up to $0.032$ of blocks across architectures, against $0.89$ for a mutating attacker. Timing-based detection of a \emph{passive} attacker is not robust: the configuration that detects false-positives on benign timing variance, while the false-positive-free configuration is blind to it, and neither reproduces across identical runs. Therefore, the checksum audits active attacks that do not decouple the audited computation from the attacked one. This qualifier is not a formality, since an attacker answering on an untouched copy while extracting from a replica passes at a TPR of $0.00$, bit-identical to an honest client. No protocol verifying only a client's reported computation can do better. Closing that gap needs evidence about total resource consumption -- attested execution, or a proof of training~\cite{tmtfl2025}. Detecting passive inference robustly remains open.

\section*{AI use disclosure}
\label{sec:disclosure}
Generative AI tools were used to help set up experiments as well as for editorial purposes in this manuscript. The authors reviewed, verified, and approved all outputs to ensure accuracy and originality, and remain responsible for the paper.


\begin{thebibliography}{99}
\bibitem{mcmahan2017} H. B. McMahan, E. Moore, D. Ramage, S. Hampson, and B. Ag\"uera y Arcas, ``Communication-efficient learning of deep networks from decentralized data,'' in \emph{AISTATS}, 2017.
\bibitem{shokri2017} R. Shokri, M. Stronati, C. Song, and V. Shmatikov, ``Membership inference attacks against machine learning models,'' in \emph{IEEE S\&P}, 2017.
\bibitem{nasr2019} M. Nasr, R. Shokri, and A. Houmansadr, ``Comprehensive privacy analysis of deep learning: Passive and active white-box inference attacks against centralized and federated learning,'' in \emph{IEEE S\&P}, 2019.
\bibitem{melis2019} L. Melis, C. Song, E. De Cristofaro, and V. Shmatikov, ``Exploiting unintended feature leakage in collaborative learning,'' in \emph{IEEE S\&P}, 2019.
\bibitem{hitaj2017} B. Hitaj, G. Ateniese, and F. P\'erez-Cruz, ``Deep models under the GAN: Information leakage from collaborative deep learning,'' in \emph{ACM CCS}, 2017.
\bibitem{abadi2016} M. Abadi, A. Chu, I. Goodfellow, H. B. McMahan, I. Mironov, K. Talwar, and L. Zhang, ``Deep learning with differential privacy,'' in \emph{ACM CCS}, 2016.
\bibitem{dwork2006} C. Dwork, F. McSherry, K. Nissim, and A. Smith, ``Calibrating noise to sensitivity in private data analysis,'' in \emph{Theory of Cryptography Conference (TCC)}, 2006.
\bibitem{blanchard2017} P. Blanchard, E. M. El Mhamdi, R. Guerraoui, and J. Stainer, ``Machine learning with adversaries: Byzantine tolerant gradient descent,'' in \emph{NeurIPS}, 2017.
\bibitem{yin2018} D. Yin, Y. Chen, K. Ramchandran, and P. Bartlett, ``Byzantine-robust distributed learning: Towards optimal statistical rates,'' in \emph{ICML}, 2018.
\bibitem{shejwalkar2021} V. Shejwalkar and A. Houmansadr, ``Manipulating the Byzantine: Optimizing model poisoning attacks and defenses for federated learning,'' in \emph{NDSS}, 2021.
\bibitem{zhang2023} Y. Zhang \emph{et al.}, ``AgrEvader: Poisoning membership inference against Byzantine-robust federated learning,'' in \emph{The Web Conference}, 2023.
\bibitem{yeom2018} S. Yeom, I. Giacomelli, M. Fredrikson, and S. Jha, ``Privacy risk in machine learning: Analyzing the connection to overfitting,'' in \emph{IEEE CSF}, 2018.
\bibitem{opacus} A. Yousefpour \emph{et al.}, ``Opacus: User-friendly differential privacy library in PyTorch,'' arXiv:2109.12298, 2021.
\bibitem{carlini2022} N. Carlini, S. Chien, M. Nasr, S. Song, A. Terzis, and F. Tram\`er, ``Membership inference attacks from first principles,'' in \emph{IEEE S\&P}, 2022.
\bibitem{zhu2019} L. Zhu, Z. Liu, and S. Han, ``Deep leakage from gradients,'' in \emph{NeurIPS}, 2019.
\bibitem{fedmia2025} G. Zhu, D. Li, H. Gu, Y. Yao, L. Fan, and Y. Han, ``FedMIA: An effective membership inference attack exploiting `all for one' principle in federated learning,'' in \emph{IEEE/CVF CVPR}, 2025.
\bibitem{tmtfl2025} Y. Wang \emph{et al.}, ``TMT-FL: Enabling trustworthy model training of federated learning with malicious participants,'' \emph{IEEE Trans. Dependable and Secure Computing}, 2025.
\bibitem{bai2025survey} L. Bai, H. Hu, Q. Ye, H. Li, L. Wang, and J. Xu, ``Membership inference attacks and defenses in federated learning: A survey,'' \emph{ACM Computing Surveys}, 2025.
\end{thebibliography}
\end{document}